\documentclass[conference]{IEEEtran}
\usepackage{preamble}
\usepackage{subcaption}

\begin{document}

\title{Scanning the Internet for ROS:  A View of Security in Robotics Research}

\author{\authorblockN{Nicholas DeMarinis, Stefanie Tellex, Vasileios Kemerlis, George Konidaris, Rodrigo Fonseca}
\authorblockA{Computer Science Department\\
Brown University\\
Providence, Rhode Island, 02912\\
Email: \{ndemarin, stefie10, vpk, gdk, rfonseca\}@cs.brown.edu}}

\maketitle

\begin{abstract}
  Because robots can directly perceive and affect the physical world, security
  issues take on particular importance. In this paper, we describe the results
  of our work on scanning the entire IPv4 address space of the Internet for
  instances of the Robot Operating System (ROS), a widely used robotics
  platform for research. Our results identified that a number of hosts
  supporting ROS are exposed to the public Internet, thereby allowing anyone to
  access robotic sensors and actuators. As a proof of concept, and with
  consent, we were able to read image sensor information and move the robot of
  a research group in a US university. This paper gives an overview of our
  findings, including the geographic distribution of publicly-accessible
  platforms, the sorts of sensor and actuator data that is available, as well
  as the different kinds of robots and sensors that our scan uncovered.
  Additionally, we offer recommendations on best practices to mitigate these
  security issues in the future.
\end{abstract}

\IEEEpeerreviewmaketitle

\section{Introduction}
\label{sec:introduction}

Security is particularly important in robotics platforms.  A robot can
sense the physical world using sensors, or directly change the
physical world with its actuators.  Thus, a robot can leak sensitive
information about its environment if accessed by an unauthorized
party, or even cause physical harm if operated unsafely.  Existing
work has assessed the state of industrial robot security and found a
number of
vulnerabilities~\cite{quarta2017experimental,maggi2017rogue}. However,
we are unaware of any studies that gauge the state of security in
robotics research.

To address this problem we conducted several scans of the whole IPv4
address space, in order to identify un\-protected Robot Operating
System (ROS) hosts~\citep{quigley09}, which are widely used in
robotics research.  Like many research platforms, the ROS designers
made a conscious decision to exclude security mechanisms because they
did not have a clear model of security threats and were not security
experts themselves. The ROS master node trusts all nodes that connect
to it, and thus should not be exposed to the public Internet.
Nonetheless, our scans identified over 100 publicly-accessible hosts
that are running a ROS master node. Of the nodes we found, a number of
them are connected to simulators, such as
Gazebo~\citep{koenig2004design}, while others appear to be real robots
capable of being remotely moved in ways dangerous both to the robot
and the objects around it. We present both a quantitative and
qualitative overview of our findings.

Quantitatively, we assessed the number of topics that appear to be sensors and
actuators of various types. We noticed a roughly Zipfian distribution, with a
few very common types, but a long tail of one-off sensors and actuators. We
also observed that many robots seem to come online for a relatively short
period (hours or days) and then go offline again. They are not generally left
accessible for an extended period of time, although we found some that were.

\begin{figure}
\includegraphics[width=\columnwidth]{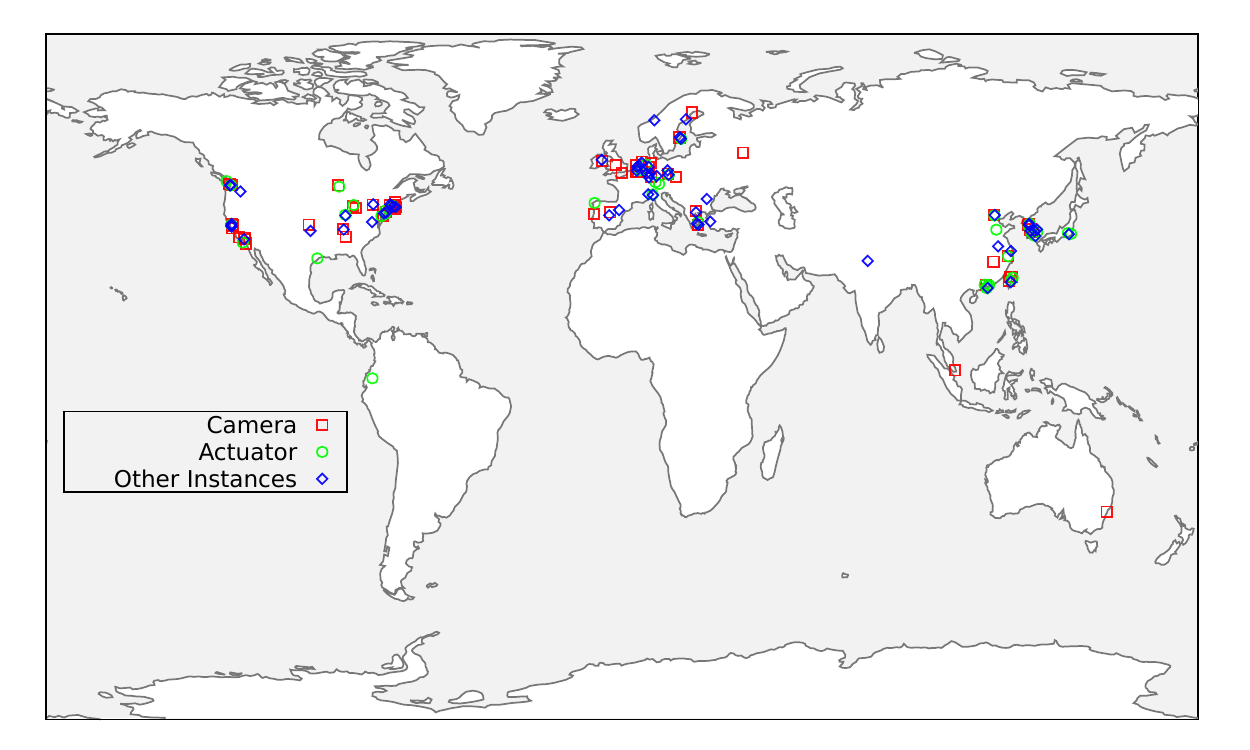}
\caption{Approximate locations (slightly jittered to show multiple
  points) of identified ROS masters across all scans. Red indicates a
  host that showed evidence of publishing camera information.  Blue
  indicates a host that showed evidence of a robot that could be
  actuated. Other hosts are in green.}
\vspace{-7mm}
\end{figure}

Qualitatively, we present case studies of several robots that we found, in
order to give a flavor of our results. We also present a proof-of-concept
takeover of one of the robots (with the consent of the robot owner), in order
to show that an open ROS master indicates a robot whose sensors can be remotely
accessed, and whose actuators can be remotely controlled.

This scan was eye-opening for us as well. We found two of our own
robots as part of the scan, one Baxter~\cite{baxter} robot and one
drone. Neither was intentionally made available on the public
Internet, and both have the potential to cause physical harm if used
inappropriately.

Our goal is not to single out any researchers or robot platforms, but
to promote security as an important consideration---not just in
production systems, but in research settings as well.  Instead, our
aim is to provide information about a concerning situation and
guidance about how the robotics community can improve their
security. \emph{Note that before the release of this work, we
  have reached out to the owners of all affected robots and provided
  them with a summary of our findings.}

\section{Related Work}
\label{sec:related-work}

Anecdotally, we are aware of a number of compromised robots. 
As one example, we have been aware that Baxter~\cite{baxterssh} robots contain a
number of security problems: besides running a ROS master on port 11311 (TCP),
the robot also runs an SSH server with a known default username and
password~\cite{baxterssh} that cannot be changed by the
owner~\cite{baxtermailinglist}. While the SSH account does not directly give
administrative access, it is still a significant security risk, as it can be
used as a stepping stone in a multistage attack. That said, we are unaware of
any specific Baxters that were compromised as a result of this issue.

OSRF is developing a set of security enhancement to ROS, termed Secure Robot
Operating System (SROS)~\citep{sros}. As of the current writing, its website
declares that this effort is ``highly experimental and must not be considered
production-grade.'' It includes TLS~\cite{rfc5246} support for all socket
transport within ROS, certificates for chains of trust, as well as namespace
and node restrictions. This effort would certainly mitigate the security issues
we have discovered; \eg a properly configured chain of trust would only allow
authorized nodes to publish commands to topics that cause the robot to move.
However, given the current state of the software, it is not yet in wide use.
\citet{dieber2016application} propose a new architecture for an authentication
and authorization services that can limit access to ROS services.
Additionally, we suspect that a more systemic solution should be deployed in
addition to these methods: the port 11311 should be filtered, by most hosts,
rather than configuring a complicated authentication/authorization
infrastructure.

Rosbridge~\cite{crick12} provides a WebSocket interface to ROS and a server to
allow web applications interact with ROS nodes. Rosbridge is often used as part
of Robot Web Tools~\cite{toris15} to make robots accessible from the Internet.
To facilitate this, Rosbridge provides TLS support for WebSocket connections,
access control(s) to limit the topics available to clients, and an
authorization mechanism to restrict API calls~\cite{toris15}. Unfortunately,
all these features are optional and disabled by default.

\citet{denning2009spotlight} studied the security and privacy risks of home
robots. They investigated three specific household robots in 2008, uncovered a
number of vulnerabilities, and surveyed implications for the
future. \citet{mcclean2013preliminary} provide an overview of various security
vulnerabilities, discovered in ROS, after setting up a honeypot at DEFFCON-20
(2012). They report several issues, including plaintext communications,
unprotected TCP ports, and unencrypted data storage. \citet{cerrudo2017hacking}
identified vulnerabilities in several home and research robotics platforms in
2017, demonstrating weak or non-existent authentication procedures that allow an
attacker to control robots, perform firmware updates, or gain access to sensor
data.  Our contribution is to identify instances of of likely-exploitable
configurations on the public Internet.

\citet{quarta2017experimental} and \citet{maggi2017rogue} conduct a
security assessment on an industrial robot controller, combined with a
practical exploit of an arm. They surveyed domain experts from both
academia and industry, and found that 30\% had robots accessible from
the Internet, while 76\% had never performed a professional
cybersecurity assessment. They also used public Internet search
engines, like Shodan and ZoomEye, to look for FTP servers matching
industrial devices, identifying 28 robots and thousands of
``industrial routers'' that enable remote access to devices. Our
results, instead quantitatively assess the number of ROS robots we
found accessible to the public Internet.

The Shodan search engine~\citep{shodan} continually scans the Internet for
services on a number of ports, and allows public searches of the results.
Examples of ports scanned by Shodan include webserver ports (80, 443, 8080),
FTP (21), SSH (22), and Telnet (23). It does not currently report scans for the
default ROS master port: \ie TCP port 11311.

\section{Robot Operating System (ROS)}
\label{sec:robot-oper-syst}

The Robot Operating System (ROS)~\citep{quigley09} was introduced by Willow
Garage in 2007. ROS operates as a publish-subscribe service to distribute data
among \emph{nodes} in a system. A central master service is responsible for
tracking published and subscribed topics and provides a parameter server for
nodes to store various metadata. Nodes can publish data as \emph{topics} by
advertising to the ROS master service. Other nodes can subscribe to these
topics by querying the master, which provides the IP address and TCP port
number of any nodes publishing a given topic, allowing the subscriber to
contact the publishers directly to establish further connections.

ROS has a distributed architecture: nodes may run on the same machine as the
master, or on different machines. The ROS master API is implemented using the
XML-RPC protocol, which is built over HTTP, and typically listens on TCP port
11311~\citep{rosmaster}. Each node runs its own slave XML-RPC server that
allows the node to advertise the topics it publishes other nodes, update
parameters, get the topics the node publishes and subscribes to, and receive
parameter and publisher updates from the ROS master~\citep{rosslave}.

When subscribing to a topic, a client first asks the ROS master for all nodes
that publish the topic. Then, it connects to each node's XML-RPC server running
the slave API and calls a method to negotiate a connection. This method tells
the publishing node to open a TCP port for the subscriber node to connect to
and ``speak'' the TCPROS protocol. Note that all of this communication is
handled internally by the ROS client libraries (most often, \texttt{roscpp} or
\texttt{rospy}).

\section{Identifying Robots using ROS}
\label{sec:ident-robots-using}

We searched for ROS masters connected to the IPv4 Internet address
space by performing a scan of all public addresses (roughly 3.7
billion IPs) on TCP port 11311, the default ROS master port. Our
scans were performed using ZMap~\cite{durumeric2013zmap}, a research
tool for performing Internet-wide port scans. Port scans operate by
asynchronously sending probe packets (\texttt{TCP SYN}) to a set of
addresses to gather information about the hosts that respond.

While port scans are very common on the public Internet, conducting
Internet-wide scans poses some inherent risks. First, and foremost, the volume
of traffic sent by a scan could overwhelm destination networks. For this
reason, we chose ZMap as our scan apparatus, because it selects addresses to
scan using a pseudorandom permutation, rather than probing all addresses in
sequential order, to greatly reduce the number of packets sent to a network at
one time. Second, sending ROS commands to unknown hosts also has the potential
to cause disruption: \eg{}, if the host is running a service other than a ROS
master on port 11311. When designing our scanning framework, \emph{we made
efforts to minimize potential disruptions by using a series of
minimally-invasive probes to confirm the presence of a ROS master before
sending commands.} Specifically, our scans were conducted in four stages, each
run on successively fewer hosts:

\begin{enumerate}
	\item A \texttt{TCP SYN} scan to identify hosts accepting connections
		on the ROS master port (TCP 11311).

	\item A \texttt{TCP SYN} scan on a high-numbered, normally-closed port
		(\eg{} 58243), to rule out addresses that may respond on
		\emph{any} port to deter scanning~\cite{mirian2016internet}.

	\item An \texttt{HTTP GET /} request on the ROS master port. XML-RPC
		servers, including the ROS master, utilize HTTP and respond to
		this request with a specific error code, ruling out other
		services.

	\item A series of \emph{passive} ROS commands to collect host
		information.
\end{enumerate}

We also performed a scan for Rosbridge instances, which run on TCP port 9090
and communicate using a JSON-based WebSocket protocol. The scanning process
for Rosbridge followed similar stages, with the exception of using a different
test in stage 3 to identify WebSocket-capable HTTP servers before sending ROS
commands.

Critically, sending commands to active robots may pose a safety hazard. We
selected a minimal set of \emph{passive} commands designed to confirm that the
host was in fact running ROS, and gather data on the topics and parameters
available on each ROS instance. \emph{At no time did we attempt to modify the
state of the ROS master, or connect to any nodes,} with the exception of the
experiments discussed in Section~\ref{sec:casestudy}, which were performed with
express permission from the robot's operators. Specifically, we called {\tt
getSystemState} to retrieve the list of publishers, subscribers, and services,
and called {\tt getParamNames} to get the list of all named parameters (but not
their value). Additionally, we retrieved the value of the parameter indicating
the ROS version, as well as the {\tt robot\_description} parameter which
returns the URDF (Unified Robot Description Format) if
present~\cite{rosurdf}.

Our scans were conducted from a host located on our (university
campus) network. Each scan was performed over a period of 7 weekdays,
to ensure a low rate of probe traffic at destination networks.  In
accordance with a set of best practices outlined by the ZMap
authors~\cite{durumeric2013zmap}, we made efforts to provide
information to any users observing our scan traffic. This included
publishing a web page on our scanning host (public) IP address, with a
description of the scan and a contact email address for more
information or to request removal from further scans.  (Note that over
the course of our scanning period, we received only one
\emph{automated} request to cease probing an organization's network,
and complied with the request.) We also added a host that listens on
TCP port 11311 and logs any connections to it; the only connection we
observed was from our own scan. Thus, we have not observed any
evidence that anyone else is (yet) scanning on port 11311 to identify
publicly-accessible robots, even though, with only one server
listening, this is far from conclusive.

\section{Quantitative Results}
\label{sec:quantitative-results}

We conducted three scans on the ROS master port between December 2017 and January
2018. We refer to these scans as Master 1--3, respectively. Each ROS master
scan observed over 100 ROS instances, spanning 28 countries, with over 70\% of
the observed instances using addresses belonging to various university networks
or research institutions. We performed one scan for Rosbridge instances in
November 2017 and identified 15 total instances, with 11 instances located in
networks recognizable as cloud service providers.

\begin{table}
\caption{Identified ROS versions.}
\label{tab:result-version}
\centering
\begin{tabular}{@{}c c r r r r@{}}
\toprule
\specialcell{\textbf{ROS}
\\%
\textbf{Version}}&\textbf{Distro}&\textbf{Master 1}&\textbf{Master 2}&\textbf{Master 3}&\textbf{Rosbridge}\\%
\midrule
1.10.x&Hydro&2&1&2&0\\%
1.11.x&Indigo/Jade&60&56&44&4\\%
1.12.x&Kinetic&78&61&53&10\\%
1.13.x&Lunar&4&4&3&1\\\bottomrule
\end{tabular} \vspace{-2mm}
\end{table}

\begin{table}
\caption{Scan results summary.}
\label{tab:result-summary}
\centering
\begin{tabular}{@{}l r r r r@{}}
\toprule
\textbf{Category}&\textbf{Master 1}&\textbf{Master 2}&\textbf{Master 3}&\textbf{Rosbridge}\\%
\midrule
Identified robots&19&13&12&4\\%
Simulation only&37&32&21&2\\%
Empty ROS cores&37&29&26&0\\%
Only sensors&24&28&18&2\\%
Only actuators&2&1&3&0\\%
Only identified services&11&8&12&6\\%
Unclassified&14&11&10&1\\%
\midrule
\textbf{Total ROS Instances}&\textbf{144}&\textbf{122}&\textbf{102}&\textbf{15}\\\bottomrule
\end{tabular} \vspace{-6mm}
\end{table}

Table~\ref{tab:result-version} shows the distribution of ROS versions
for the three scans, based on the \texttt{ros\_comm} version returned
by each instance.  Almost all hosts were running either Indigo, Jade,
or Kinetic.
Table~\ref{tab:result-summary} provides a summary of our results,
organized into types based on their topic data. We define a simulator
to be a host that showed evidence of one of the robot simulator topics
listed in Section~\ref{sec:simulators}.  We define a robot as a host
that shows evidence of a sensor and an actuator, but is not a
simulator. Each type is mutually exclusive, so identified sensors,
actuators, and robots, in this table, did not show evidence of being a
simulator. These results must be taken as approximate, since
we did not actually subscribe to any of the topics, consider their
type, or verify connectivity with real hardware. However, given the
standardization of topic names, it seems likely that many of the hosts
we found were indeed running sensors, actuators, and other
software. Empty ROS cores showed only the base {\tt rosout} topics and
services. Some of these hosts additionally had parameters set (perhaps
indicating some nodes had been running previously, and later shut
down).  Unclassified nodes did not fit into any of our other
categories.

We do not combine the results of each scan to form a ``grand total''
for each type, as many hosts appeared in more than one scan and
returned different topic data each time. We discuss these observations
in more detail in Section~\ref{sec:persistent-hosts-vs}.

\begin{table*}
\caption{Topic and parameter search results.}
\label{tab:search-sensors}
\centering
	\renewcommand{\arraystretch}{1.0}
\begin{tabular}{@{}l l r r r r r r r r@{}}
\toprule
\textbf{\textbf{Category}}&\textbf{\textbf{Parameter}}&\multicolumn{2}{c}{\textbf{Master 1}}&\multicolumn{2}{c}{\textbf{Master 2}}&\multicolumn{2}{c}{\textbf{Master 3}}&\multicolumn{2}{c}{\textbf{Rosbridge}}\\%
\midrule
\multirow{25}{*}{Sensors}&\textbf{}&\textbf{Phys. HW}&\textbf{Sim./Log}&\textbf{Phys. HW}&\textbf{Sim./Log}&\textbf{Phys. HW}&\textbf{Sim./Log}&\textbf{Phys. HW}&\textbf{Sim./Log}\\%
\cmidrule(lr){3
-
4}
\cmidrule(lr){5
-
6}
\cmidrule(lr){7
-
8}
\cmidrule(l){9
-
10}
&Camera&29&22&29&11&22&11&5&1\\%
&\hspace*{0.5em}
Camera + Depth&13&15&13&6&9&8&2&1\\%
&\hspace*{0.5em}
Camera + RGB&12&7&6&5&9&5&1&1\\%
&\hspace*{0.5em}
Camera + Stereo&1&3&1&4&&3&&\\%
&Kinect&3&2&3&2&2&2&&\\%
&IMU&14&16&9&16&6&11&2&\\%
&Gyro&&&&&&&&\\%
&Lidar&12&13&5&9&2&11&3&2\\%
&Motion Capture&4&2&3&1&&&&\\%
&Compass&&3&&2&&1&&\\%
&Odometry&8&12&6&14&5&11&3&2\\%
&Pressure&1&3&1&2&1&1&&\\%
&Contact&4&3&2&3&2&4&&\\%
&\texttt{biotac}&1&&&&&&&\\%
&Velodyne&4&5&4&2&3&1&&\\%
&\texttt{point\_cloud}&1&4&&1&3&1&&1\\%
&Force&&1&&&&&&\\%
&Radar&1&2&1&1&1&1&&\\%
&Geolocation&4&9&6&8&3&3&&\\%
&Audio&&&1&&&&&\\%
&Temperature&&2&2&1&1&&&\\%
&Battery Monitor&4&3&3&2&2&1&5&\\%
&Printhead status&&&&&1&&&\\%
&Joystick&5&2&3&9&2&3&3&\\%
\midrule
\multirow{15}{*}{Actuators}&\textbf{}&\textbf{Phys. HW}&\textbf{Sim./Log}&\textbf{Phys. HW}&\textbf{Sim./Log}&\textbf{Phys. HW}&\textbf{Sim./Log}&\textbf{Phys. HW}&\textbf{Sim./Log}\\%
\cmidrule(lr){3
-
4}
\cmidrule(lr){5
-
6}
\cmidrule(lr){7
-
8}
\cmidrule(l){9
-
10}
&Movable base&11&12&8&13&9&9&4&2\\%
&Servo&1&1&2&&&&&\\%
&Lights&1&12&1&9&1&6&&\\%
&Arm&4&6&&7&2&3&&1\\%
&Gripper&4&3&1&5&2&2&&1\\%
&Flippers&&&&5&&&&\\%
&Sound&1&&1&&2&&&\\%
&Heartbeat&2&&1&&1&&&\\%
&Voice&3&&&&&&&\\%
&MotorCommand&1&&&&&&&\\%
&\texttt{inceptor\_command}&1&&1&&1&&&\\%
&\texttt{flystate2phidgetsanalog}&&&1&&1&&&\\%
&Emergency Stop&&2&&2&&1&&\\%
&Printhead&&&&&1&&&\\%
\midrule
\multirow{7}{*}{Simulators}&\textbf{}&\textbf{Phys. HW}&\textbf{Sim./Log}&\textbf{Phys. HW}&\textbf{Sim./Log}&\textbf{Phys. HW}&\textbf{Sim./Log}&\textbf{Phys. HW}&\textbf{Sim./Log}\\%
\cmidrule(lr){3
-
4}
\cmidrule(lr){5
-
6}
\cmidrule(lr){7
-
8}
\cmidrule(l){9
-
10}
&Gazebo&&19&&18&&15&&1\\%
&Unity&&1&&1&&&&\\%
&Stageros&&1&&2&&1&&1\\%
&\texttt{torcs\_ros}&&1&&&&&&\\%
&Dreamview&&2&&1&&1&&\\%
&Playback&&3&&2&&&&\\%
\midrule
\multirow{7}{*}{Robot Types}&\textbf{}&\textbf{Phys. HW}&\textbf{Sim./Log}&\textbf{Phys. HW}&\textbf{Sim./Log}&\textbf{Phys. HW}&\textbf{Sim./Log}&\textbf{Phys. HW}&\textbf{Sim./Log}\\%
\cmidrule(lr){3
-
4}
\cmidrule(lr){5
-
6}
\cmidrule(lr){7
-
8}
\cmidrule(l){9
-
10}
&Baxter&1&&1&&1&&&\\%
&PR2&&2&&3&&2&&\\%
&WAM&1&1&&1&&&&\\%
&JACO&1&&&&&&&\\%
&Turtlebot&1&&&&1&&&\\%
&DaVinci&&&1&&&&&\\%
\midrule
\multirow{13}{*}{Libraries}&\textbf{}&\textbf{Phys. HW}&\textbf{Sim./Log}&\textbf{Phys. HW}&\textbf{Sim./Log}&\textbf{Phys. HW}&\textbf{Sim./Log}&\textbf{Phys. HW}&\textbf{Sim./Log}\\%
\cmidrule(lr){3
-
4}
\cmidrule(lr){5
-
6}
\cmidrule(lr){7
-
8}
\cmidrule(l){9
-
10}
&Rosbridge&7&3&8&3&9&2&12&2\\%
&RViz&27&15&19&7&15&1&&\\%
&MoveIt!&1&4&&2&1&&&\\%
&OpenRAVE&1&&1&&1&&&\\%
&Transform Library (tf)&39&28&32&17&26&15&4&2\\%
&Fiducial Libraries&1&2&1&2&&1&&\\%
&ROS Tutorials&&1&&1&&1&&\\%
&\texttt{master\_discovery}&2&&3&&2&&1&\\%
&\texttt{master\_sync}&2&&3&&2&&1&\\%
&\texttt{robot\_position}&1&&&&1&&&\\%
&\texttt{robot\_position}&1&&&&1&&&\\%
&\texttt{web\_video\_server}&2&1&2&1&1&2&2&1\\\bottomrule
\end{tabular} \end{table*}

Table~\ref{tab:search-sensors} shows the number of hosts for every type of
sensor, actuator, and service, for each scan. Each host may appear in more than
one row of this table, for example, if it contained evidence of a camera, an
IMU (Inertial Measurement Unit), or a joystick. We separate results into hosts
that showed evidence of being a simulator vs. all others. This way, we can
separate hosts that showed evidence of exposing physical sensors compared to
hosts that are likely only exposing simulated sensors and actuators. We
inferred information about robots from the {\tt robot\_description} parameter,
as well as some canonical topic names. 
The most common type of robot in our search was the
Turtlebot~\cite{turtlebot}.
We also found Baxter robots~\cite{baxter}, WAM arms~\cite{wamarm}, the
Da Vinci research kit~\citep{kazanzides2014open}, drones, vehicles,
and one flying insect.

\subsection{Identifying topics}
\label{sec:identifying-topics}

We classified the number of sensors and actuators in different categories. 
Because we wanted to limit the load we placed on discovered host, we
only retrieved the names of topics, but not their type. While this
fact does limit our knowledge, because we cannot search, for example,
for all {\tt sensor\_msgs::Image} topics, we can still obtain
substantial evidence from the topic names.
We did not subscribe or publish to any topics, because we did not want
to view any sensor data or actuate any robots without permission. As a
result, we cannot state with certainty whether we have found active
sensors or actuators. However, the list of topics provides evidence of
what is likely to be available to an attacker. 

With the obtained data in a local database, we constructed queries for
various topics of interest in robotics, and divide our results into
sensors and actuators.  We also looked for common libraries, and
evidence of use of simulators.
In constructing queries, we manually examined the topic list in order to
identify topics that map to different sensors/actuators. As we found
indicators, we added these to the list of queries run on all data to classify
the object. We describe the queries below; the complete list can be found in
our supplementary attachment.

\subsection{Sensors}
\label{sec:sensors}

Sensors found in our scan included cameras, laser range finders, barometric
pressure sensors, GPS devices, tactile sensors, and compasses. While camera
topics appear to have the most standardization in terms of topic
names (\eg {\tt camera\_info} or {\tt image}), we
found other viable search terms as well. More specifically: (a)~{Cameras:} we
searched for a number of standard camera sensors,
as well as for depth cameras, by looking for {\tt depth\_registered}.
(b)~{LIDAR:} we searched for {\tt velodyne} and {\tt point\_cloud}.
(c)~{Barometric Pressure Sensors:} we found barometric pressure sensors that
appear to be used to sense height in drones; the search term {\tt baro} was used
to identify these devices. (d)~{Tactile Sensors:} we found evidence of a
specific tactile sensor, the BioTac sensor, using the topic {\tt biotac}.
(e)~{Compass Sensors:} we found several hosts that appeared to have compasses,
by using the topic {\tt compass}. (f)~{Odometry:} we found odometry sensors,
which had the fairly standard name {\tt odom} or {\tt odometry}. (g)~{Joystick
  and Keyboard:} we searched for topics with the name \texttt{joy} or
\texttt{joystick} for evidence. (h)~{Microphone:} we searched for the topic {\tt
  microphone} to identify audio sensors.

\subsection{Actuators}
\label{sec:actuators}

Actuator topic names are much more eclectic. We found some standardization in
topics for moving joints, but the majority of the topic names we observed, in
this case, were one-off. In particular: (a)~{Arms:} we found a number of
standard topics that seemed to indicate an arm that moves, including {\tt
joint\_trajectory}, {\tt trajectory\_controller}, and {\tt action\_controller}.
(b)~{Grippers:} many robots that had grippers used {\tt gripper} as a topic
name. (c)~{Playing Sound:} we found evidence that the \texttt{sound\_play}
library was used to play sounds on several hosts, and, hence we used that topic
to identify sound actuators. (d)~{Heartbeat:} a heartbeat is an indicator that
an actuator is involved because it is important to have a heartbeat if the
robot can move and therefore potentially hurt someone; we searched for the
topic name {\tt heartbeat}. (e)~{One-Off:} we found a number of one-off
actuator topics; for one manipulator robot, the topic {\tt MotorCommand} seemed
to indicate an actuator; for a drone, the topic {\tt inceptor\_command} seemed
likely to refer to a topic that sends flight commands; the topic {\tt
flystate2phidgetsanalog} seems to send voltages
for controlling a faux-fruit fly~\citep{kinefly}.  

\subsection{Simulators}
\label{sec:simulators}

We also found many ROS instances that appeared to be connected to simulated
robots. The most obvious were connected to the Gazebo
simulator~\citep{koenig2004design}. We also found groups that appeared to be
using the Unity Game Engine~\citep{engine9unity}, the game
TORCS~\citep{torcsros}, and the Stage simulator~\citep{gerkey2003player}. We
considered any topics with {\tt gazebo}, {\tt unity}, or {\tt torcs\_ros} to be
evidence of using one of these simulators. Additionally, if the host had the
parameter {\tt use\_sim\_time} or \texttt{fake}, indicating use of simulated
inputs, we considered it to be a simulator as well.

Running a simulator with an open ROS master does not pose the same physical risk
as a real robot.  However, given the complexity of a ROS system, and the ability
to connect to other nodes, in many programming languages, it seems likely that
this configuration still poses a threat: the machine can be compromised through
a remote exploit, such as a buffer overrun in a ROS node.

\subsection{Common Libraries}
\label{sec:common-libraries}

We also report evidence of use of the most common libraries, including
MoveIt!~\citep{chitta2012moveit}, Gazebo~\citep{koenig2004design} and
others.

\paragraph{Rosbridge}
Rosbridge~\citep{crick12} provides a JSON-based interface to ROS using
WebSockets, allowing client libraries to interface with ROS from a
single connection. We searched for {\tt rosbridge} to identify hosts running
this node.

\paragraph{MoveIt!}
MoveIt!~\citep{chitta2012moveit} is a library that performs motion planning
primarily for robot arms. We searched for the topic {\tt move\_group} for
evidence that this library was being run. We identified many MoveIt! instances
that were not connected to real robots, but some were indeed connected to real
arms.

\paragraph{Fiducial Libraries}
It is common to use fiducial marker libraries, such as
AprilTags~\citep{olson2011apriltag} or AR Tags~\citep{fiala2005artag}, to mark
objects in the robot's environment. We searched for hosts using these libraries
with the topics {\tt apriltag} and {\tt ar\_track\_alvar}.

\subsection{Persistent Hosts vs. Temporary Hosts}
\label{sec:persistent-hosts-vs}

We hypothesized that many ROS instances are not always online, and instead are
only available (and visible to our scans) during intermittent periods of usage.
Since our scans took place over a period of 7 days, intermittent hosts may not
have been detected, if they were not available at the time their IP address was
scanned. Conversely, a host may have been detected multiple times, in a single
scan, if its IP address changed during the scan period. 
We examined our results to find hosts that appeared in multiple scans and those
appearing in only one scan. Comparing, however, host responses between scans is
non-trivial. A single ROS instance may use different IP addresses, or DNS
names, over time due to dynamic IP allocations (via DHCP) or physical movement
between networks, and may provide different topic and parameter data based on
the host's usage at the time of scanning.

We were able to match a number of hosts between scans using their local machine
hostnames, which are included in the names of parameters added to the parameter
server when starting a ROS instance or launching a node. For example, the
parameter \texttt{/roslaunch/uris/host\_potato\_\_46636} refers to a physical
machine named ``potato.'' Although a host may contain multiple
\texttt{roslaunch} parameters, if nodes are run from other machines, we found
that 90\% of the hosts scanned contained a single machine hostname
When a machine name was
not available, we also considered hosts with the same IP addresses as matches,
if they contained a percentage of similar topics. We acknowledge that using
machine hostnames to uniquely identify hosts is not definitive: \eg 
two hosts may share the same machine name, perhaps if their operating systems
were from a cloned image; we observed 5 instances of a machine name appearing
more than once, in a single scan, which either indicates two hosts with the same
machine name, or one host using different IPs.

Using these empirical methods, we found that 41 hosts were present in
all three ROS master scans, 48 hosts were found in two scans, with the
remaining hosts only observed once. Of the hosts that were matched
between scans, 25 hosts were grouped into different categories
(cf. Table~\ref{tab:result-summary}) based on their topic data at the
time each scan was performed. This demonstrates how a single ROS
instance may have different usage patterns over time, depending on the
experiments being performed by the user. While this poses a challenge
to classification, it also highlights a further privacy risk, as the
topic data made available by an open ROS instance can demonstrate a
researcher's usage patterns. To provide one example, we found one
machine name that appeared three different times in the scan with
three distinct IPs: one IP with a DNS name of a research lab at
a university, which showed topics indicating a simulator; one IP
mapped to another network in the same university, which also displayed
simulator topics; and another IP that mapped to a
consumer ISP located near the university, which displayed no topics,
indicating the ROS core was empty.

\section{Case Studies}
\label{sec:casestudy}

We present a few case studies of specific hosts we discovered, to give
a sense of the nature of robots we found, and the potential damage
that could occur if they were compromised.

\subsection{Flying Insect}
\label{sec:flying-insect}

We found what appears to be a tethered winged insect. This system appears to be
running the Kinefly node~\citep{kinefly}, which extracts kinematic variables
from a living insect including the head, abdomen, left wing and right wing. The
goal is to provide a real-time estimate of the fly's position and orientation,
in order to study fruit flies and other insects~\citep{agrawal2014relative}.
Kinefly appears to send voltages to servos that move a magnet, which causes the
dummy fly to move. It seems unlikely that by moving this robot we could have
hurt anyone, due to its small size. However, we could have certainty disrupted
any ongoing experiment, and possibly harmed the real fruit flies interacting
with the robot fly.

\subsection{Baxter}
\label{sec:baxter-1}

We found one Baxter robots in our scan, which belonged to the authors. We had
set up the robot behind a router with NAT~\cite{rfc2663} configured. However, we
wanted to enable remote access as part of our research, and hence we configured
the router to forward TCP port 11311 to the internal Baxter machine. However, we
did not disable NAT traversal past the end of our research project, and,
therefore, our robot was publicly-accessible for several months.

We took advantage of this opportunity to perform a pen test on our own robot.
We were able to run XML-RPC calls on the ROS master to obtain listed topics and
nodes, as well as modify parameter (\texttt{rosparam}) values. This access
allows us to put the robot in a dangerous state by changing the gains on the
PID controller for Baxter's arm---incorrectly tuned PIDs could cause the arm to
oscillate unexpectedly. We were unable to subscribe to topics on this robot as
it was located behind a NAT, which blocks incoming connections. A subscriber
must connect to two other TCP ports, when subscribing to a topic (one for the
XML-RPC server, and one for the TCPROS connection), and because these ports
were not forwarded through the NAT, we could not establish a connection.
Luckily, we were unable to view sensor information from the camera, or images,
from a computer outside our network.

However, because subscribers connect to publishers, we were able to publish
topics from our remote machine. We verified that we could change the image on
the robot's face by publishing a message, as well as publish joint angles.
Potentially, we could have disabled safe mode on the robot by publishing to a
topic at a high rate, as described by the Baxter
SDK~\cite{baxtersetup}.
Certainly this access would put operators of the robot at risk, since an
attacker could move the robot unexpectedly, as well as change parameters to
put the robot in a dangerous state, despite not having access to sensor
data.

\subsection{Mobile-Manipulator}
\label{sec:mobile-manipulator}

One robot found in our scan was the mobile manipulator robot Herb2, at the
University of Washington. This robot was running topics that indicated the
presence of a multisense RGB-D sensor, a node that appears to perform speech
production/generation, as well as a service for moving its neck. The robot also
had parameters indicating the presence of controllers for arms, but the arm
nodes did not appear to be running at the time of our scan.

After contacting the robot's operators, we coordinated with them to
perform a penetration test on it. They informed us that the robot was
publicly-accessible during the period of our scan, because it coincided with a
critical demonstration.

We set the {\tt ROS\_MASTER\_URI} environment variable (of our local ROS
node/host) to the URL of the robot, and {\tt ROS\_IP} to the IPv4 address of
our host. Note that for two-way communication, all nodes in ROS must be able to
connect to servers on all other nodes---the ROS master dynamically assigns a
(unique) port to each publisher. Thus, our host had to be
publicly-accessible. The next problem was that the {\tt ROS\_HOSTNAME}
variable, of the nodes in the robot's network, was set to the (local) name {\tt
herb2}, instead of the fully qualified domain name.  We fixed this problem by adding {\tt herb2} to our local {\tt
/etc/hosts} file with the correct (remote) IP address. This fix required no
changes to the robot or extra information from the remote site.

\begin{figure}
	\subcaptionbox{UW's Robot.}{\includegraphics[height=1.25in]{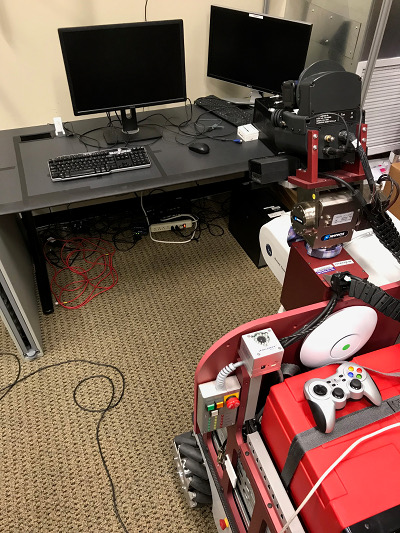}}~
	\subcaptionbox{Image obtained from the robot's camera.}{\includegraphics[height=1.25in]{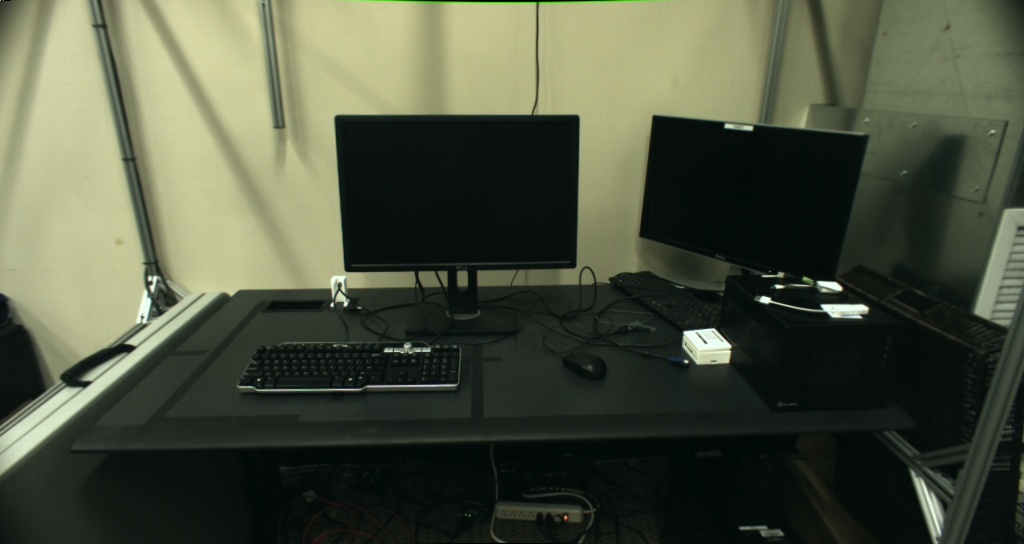}}
	\subcaptionbox{\label{fig:rviz}View from RVIZ showing 3D point cloud and TF tree.}{\includegraphics[width=1\linewidth]{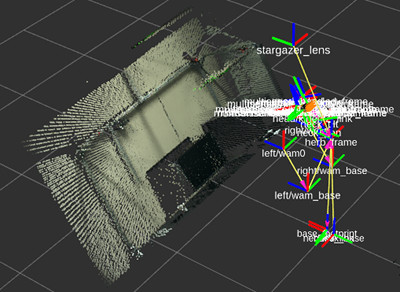}}
\caption{Images obtained from the robot's camera.\label{fig:hackcam}}
\vspace{-5mm}
\end{figure}

At this point we were able to use the camera of the ROS master, as shown in
Figure~\ref{fig:hackcam}. We could view images at 1.5Hz. Note that camera feed
can be obtained without the knowledge, or consent, of the robot owner or
operator (although we did have knowledge and consent in this case), potentially
violating the privacy of people working with the robot. Additionally, an
attacker could perform reconnaissance by viewing sensor information: \eg run a
face tracker or a movement sensor on the images, in order to determine dangerous
times to move the robot. We were also able to view joint angles, obtained from
the TF tree, as well as a 3D point cloud, visible in Figure~\ref{fig:rviz}.

We tried to actuate the robot by playing sounds and moving its neck. The robot
had a service for performing speech synthesis, and the lab had released the
code for that service on GitHub~\cite{talker}. As a result, we were able to
play sounds on the robot by installing and building this package, and writing a
\texttt{rospy} script to send the action. Many ROS topics and services have
standard names and types, like {\tt ros\_control}~\citep{roscontrol}, which
could be accessed in this way.

Finally, we tried to move the robot. The robot had a service available for
moving its neck. However this service was developed specifically for this
hardware and not released publicly by the lab. As a result, although we could
see that the service existed, we did not know the format of the {\tt srv} file
(\ie the file used by ROS for describing service types) to specify correct
calls. ROS does not allow a service call unless the MD5 checksum of the {\tt
srv} (or {\tt msg}, {\tt action}) file matches between the two nodes.
Additionally, it does not display information about the differences between the
two files when the MD5 checksum fails.

MD5 is not a (cryptographically) secure hash~\cite{md5cert}, but it can be used
to verify data integrity. It may be possible to reconstruct the \texttt{srv} (or
\texttt{msg}) file from the provided MD5 hash that is printed when the initial
attempt fails, and partial knowledge of the format.  We decided to leave the
investigation for this reconstruction for the future.  Instead, we asked the lab
to provide only the {\tt srv} file used for the service to move the neck. We
then constructed a ROS package to build the interface to the service and used
{\tt rosservice} to call it and successfully actuate the neck.  (The video
attachment documents our penetration effort, with both a screen capture from the
remote attacking computer as well as a video from the scene with the robot.)
Notably, the lab informed us that we could have damaged the robot by actuating
the neck to certain out-of-bounds values. A pan/tilt camera arm probably would
not injure a person unless their hands happened to be at a pinch point.

\section{Recommendations for Improved Security}
\label{sec:recomm-impr-secur}

We have contacted the owner(s) of the hosts identified as ROS
instances to notify them of our findings and provide recommendation
for how users can fix their security.

\parahead{Detecting exposure}
At minimum, we recommend that users inspect their network exposure using tools
such as {\tt nmap}~\cite{lyon2009nmap} and {\tt nc}~\cite{netcat}.
A command such as {\tt nmap 192.168.1.10 -p 11311} (changing the address
accordingly) will scan an address, or range of addresses, for open ROS master
ports. To correctly interpret results, it is important to consider the
network's structure and any existing security mechanisms: \eg performing the
scan from another host on the same organizational network may yield different
results than an outside network. We advise researchers to consult any
organizational policies associated with conducting scans, and potentially
coordinate with their network administrators, before scanning a wide range of
addresses.

\parahead{Firewall}
One way to secure ROS instances against unauthorized access is to use a
firewall to prevent exposure of ROS services to the public Internet. We
recommend that ROS users define a \emph{trust zone} on a network where ROS
traffic is permitted to perform tasks, and take steps to restrict network
access to ROS hosts outside this segment.

A common---but insufficient---practice to create a small private network is to
use a common wireless router, or other device, to provide Internet access to a
private address range (\eg 192.168.*.*) using NAT. By design, NAT blocks
incoming traffic that does not pertain to a connection initiated by a host on
the internal network, mimicking the operation of a firewall. Indeed, a ROS host
behind NAT would not have been visible to our scans, unless it explicitly
contained a rule to forward traffic to port 11311. However, \emph{using NAT
alone does not provide a comprehensive solution to prevent exposure of ROS
traffic to the Internet}. In general, NAT is not a security apparatus: an
internal host that makes external connections can still leak data, and NAT
implementations can be misconfigured or, in some cases, exploited to open ports
for outside access~\cite{davis2016myth,baugher2011home,garcia2011universal}.
As one example, our Baxter robot was found in our scan results because the ROS
master port for its lab was forwarded to its address.

While NAT provides a first step to blocking external traffic, we recommend
using a more complete firewall solution, possibly in coordination with existing
IT services. At minimum, traffic on the ROS master port should not be
permitted from outside hosts. ROS traffic on other services is difficult to
block, since port numbers are allocated dynamically by the ROS master---an
application-specific, stateful firewall can, however, be helpful in this case.
For client machines using ROS, we suggest using an OS-level firewall (\eg,
Netfilter~\cite{iptables}) to restrict incoming traffic on all ports except for
trusted networks. This is especially important for ROS users on laptops, or
other mobile machines, which may operate from both trusted and untrusted
networks.

In some cases, it may be necessary for a robot to be operated outside a trusted
network. It was for this reason that our Baxter was exposed to the Internet: we
had the explicit goal to operate it from an off-site location. In such cases, a
firewall provides a (far) more robust solution than NAT, as it can restrict
access based on explicit policies (\ie allow/deny access from/to
specifically-defined hosts or networks).

\parahead{Isolation}
An alternative, but less flexible, solution could involve placing ROS hosts on
an isolated network, without access to the Internet, which eliminates the
challenge of configuring firewall policies. To permit certain traffic, an HTTP
proxy (\eg Squid~\cite{squid}) can relay traffic for web browsing and critical
OS updates.

\parahead{VPN} Remote access can be provided using a virtual private
network (VPN): the VPN securely makes a remote host look as if it is a
local one, enabling the robot to be accessed remotely. If available, a
VPN provides a comprehensive way to permit only authorized hosts to
access a ROS master. While considerably simpler to configure, SSH/SSL
tunnels are generally insufficient for granting remote access, since a
ROS master may direct a remote client to connect to many different IP
addresses or ports to send or receive topic data.  VPNs are considered
the best practice for securing a ROS system by the Open Source
Robotics foundation and were configured on every PR2 that shipped.

\parahead{Rosbridge}
Rosbridge is an alternative mechanism for interacting with a ROS master that
provides certain security benefits. At minimum, Rosbridge acts as a proxy for
the ROS master API, to serve all traffic over a single port, simplifying the
challenge of making non-local connections.  Rosbridge also allows a user to
specify a list of \emph{protected} topics that will not be served by the
API. Further, it provides an optional authorization mechanism using MACs to
restrict API calls to hosts that have obtained an authorization key from an
external server~\cite{toris2014message}. While this requires users to configure
their own authentication server, it provides a way to define flexible, arbitrary
policies for access control.

\parahead{Future mechanisms}
Further extensions to ROS, such as SROS and (the emerging) ROS2~\cite{ros2},
may also help to mitigate this issue by more tightly integrating security
mechanisms into the robot architecture. Notably, ROS2's adoption of a
middleware interface for data exchange may offer a flexible way to build
appropriate defenses for a given application. Longer-term, it is important to
think about security concerns in any network, especially those running ROS. We
urge researchers using ROS to consider how the network is utilized in their
environment and take steps to protect their infrastructure.

\section{Conclusion} 
\label{sec:conclusion}

Though a few unsecured robots might not seem like a critical issue, our study
has shown that a number of research robots is accessible and controllable from
the public Internet. It is likely these robots can be remotely actuated in ways
dangerous to both the robot and the human operators. Additionally, the robot's
sensors can be viewed, which is a threat to privacy. Remote actuation has also
the ability to allow an attacker to inject subtle bugs or strange behavior into
the robot.

It is important to develop mechanisms for enabling us to use these powerful
tools while also securing them so that they cannot be accessed by bad actors.
In the future, we hope that Shodan will scan port 11311 to aid in identifying
open robots so that we can protect them. We also hope that more robosticists
will be aware of the security issues involved as a result of our work, and take
steps to prevent their robots from being publicly-accessible.

Additionally, we are eager to explore other popular platforms, such as
YARP~\citep{metta2006yarp} and LCM~\citep{huang2010lcm}. These scans may
identify additional robots that should be monitored for intrusion and
additional vulnerabilities.

As robots move out of the lab and into industrial and home settings, the number
of units that could be subverted is bound to increase manifold. Recent attacks
on the Internet infrastructure from home devices serve as a clear warning. It
will be even more important to provide adequate security for devices that can
change not just the virtual world, but the physical world as well.

\section{Acknowledgements}

We would like to thank Sidd Srinivasa and Youngsun Kim for their
help with our proof-of-concept ``attack.''

\balance
\bibliographystyle{plainnat}
\bibliography{main}

\end{document}